\RequirePackage{ifpdf}
\ifpdf 
\documentclass[pdftex]{sigma}
\else
\documentclass{sigma}
\fi

\begin{document}


\renewcommand{\thefootnote}{$\star$}

\renewcommand{\PaperNumber}{015}

\FirstPageHeading

\ShortArticleName{Harmonic Superf\/ields in ${\cal N}=4$ Supersymmetric Quantum Mechanics}

\ArticleName{Harmonic Superf\/ields in $\boldsymbol{{\cal N} = 4}$ Supersymmetric\\  Quantum Mechanics\footnote{This
paper is a contribution to the Proceedings of the Workshop ``Supersymmetric Quantum Mechanics and Spectral Design'' (July 18--30, 2010, Benasque, Spain). The full collection
is available at
\href{http://www.emis.de/journals/SIGMA/SUSYQM2010.html}{http://www.emis.de/journals/SIGMA/SUSYQM2010.html}}}

\Author{Evgeny A. IVANOV}

\AuthorNameForHeading{E.A.~Ivanov}

\Address{Bogoliubov Laboratory of Theoretical Physics,
JINR, 141980, Dubna, Moscow Region, Russia}
\Email{\href{mailto:eivanov@theor.jinr.ru}{eivanov@theor.jinr.ru}}

\ArticleDates{Received December 20, 2010, in f\/inal form February 03, 2011;  Published online February 11, 2011}

\Abstract{This is a brief survey of applications of the harmonic
superspace methods to the models of ${\cal N}=4$ supersymmetric
quantum mechanics (SQM). The main focus is on a~recent progress in
constructing SQM models with couplings to the background non-Abelian
gauge f\/ields. Besides reviewing and systemizing the relevant
results, we present some new examples and make clarifying comments.}


\Classification{81T60; 81Q60}

\renewcommand{\thefootnote}{\arabic{footnote}}
\setcounter{footnote}{0}

\section{Introduction}
Supersymmetric quantum mechanics (SQM) \cite{Witt} is the simplest (${d=1}$) supersymmetric theory. It has
plenty of applications in various domains. Some salient features  of the SQM models and their uses were already
discussed during this Benasque meeting. Let us recall some other ones:
\begin{itemize}\itemsep=0pt
\item
SQM models are capable to catch characteristic properties of the higher-dimensional supersymmetric theories
via the dimensional reduction \cite{smilR};
\item
They provide superextensions of integrable models like Calogero--Moser systems \cite{scal1,Calog} and
Landau-type models \cite{sland};
\item
An extended supersymmetry in ${d=1}$ is specif\/ic; it exhibits some features which are not shared by its $d >1$ counterparts.
These are the so-called automorphic dualities between various supermultiplets with the dif\/ferent of\/f-shell contents~\cite{automorph},
the existence of nonlinear ``cousins'' of of\/f-shell linear multiplets \cite{nonl,hss1}, and some other ones.
\end{itemize}

An ef\/f\/icient tool to deal with extended supersymmetries in ${d > 1}$ is known to be the {\it harmonic superspace} (HSS)~\cite{HSS,book}.
The HSS approach allowed to construct, for the f\/irst time, an of\/f-shell formulation of hypermultiplets in ${\cal N}=2$, $d=4$ and
${\cal N}=1$, $d=6$ supersymmetry,
as well as a formulation of ${\cal N}=4$, $d=4$ supersymmetric Yang--Mills theory with the maximal number ${\cal N}=3$ of of\/f-shell supersymmetries
(at cost of an inf\/inite number of the auxiliary f\/ields appearing in the harmonic expansions of the relevant superf\/ields).
Some further consequences of the HSS approach for the ${d > 1}$ supersymmetric theories are listed, e.g., in~\cite{book,my}.

A natural extension of the HSS approach was applying it to ${d=1}$ supersymmetric theories, i.e.\ SQM models.
An ${\cal N}=4$, $d=1$ version of the ${\cal N}=2$, $d=4$ HSS was worked out in~\cite{hss1}.
It proved to be a powerful device of the ${\cal N}\geq 4$ SQM model-building, as well as of getting new insights into the structure of
$d=1$ supersymmetries and their representations.
In particular, it allowed one to understand interrelations
between various ${\cal N}=4$ SQM models via the manifestly ${\cal N}=4$ covariant gauging procedure \cite{Deld1,Deld2,Deld3}.
As one more important application, it helped to construct new ${\cal N}=4$ superextensions of the Calogero-type models~\cite{Calog}.

The latest developments of the $d=1$ HSS approach concern applications in SQM models with the Lorentz-force type couplings to the external gauge
f\/ields, i.e.\ couplings of the form $A_m(x(t))\dot{x}^m(t)$. The major subject of this contribution is just a survey  of these new applications from
a common point of view, with some clarifying examples and further remarks.

Let us adduce some reasons why SQM models with external gauge f\/ields are of interest.

One reason is that these models supply ${d =1}$ prototypes of the $p$-branes world-volume couplings. The other one is the close
relation of such models to supersymmetric versions of the Wilson loops and Berry phase (see, e.g.,~\cite{Berry}). Also, they provide
superextensions of the Landau problem and of the quantum Hall ef\/fect (see, e.g., \cite{sland}) and give quantum-mechanical realizations
of Hopf maps (see, e.g.,~\cite{hopf}). At last, they  yield, as the particular ``extreme'' case, superextensions
of the Chern--Simons mechanics~\cite{ch}.

Our consideration will be limited to the ${\cal N}=4$ SQM models with the background gauge
f\/ield\footnote{The on-shell ${\cal N}=2$ SQM models with couplings to a non-Abelian monopole background were considered,
e.g., in~\cite{Hor}.}. Until recently, only ${\cal N}=4$ superextensions of  the couplings to {\it Abelian} background gauge f\/ields were known.
Their of\/f-shell formulation  within the ${\cal N}=4$, $d=1$ HSS setting was given in the paper \cite{hss1}.
The coupling to {\it non-Abelian} gauge backgrounds was recently constructed in
the papers \cite{KSm,IKSm,IKon} (see also \cite{BKS,KLS}). This construction essentially exploits the {\it semi-dynamical}
(or {\it spin}, or {\it isospin}) supermultiplet $({\bf 4, 4, 0})$ \cite{Calog}.
Bosonic f\/ields of the spin multiplet are described by the U(1) gauged Wess--Zumino $d=1$ action and, after quantization, yield
generators of the gauge group SU(2).
A key role is also played by the manifestly ${\cal N}=4$ supersymmetric ${d=1}$ gauging procedure worked out in the papers
\cite{Deld1,Deld2,Deld3}. It turns out that the requirement of of\/f-shell ${\cal N}=4$ supersymmetry forces
the external gauge potential to be (a)~{\it self-dual} and
(b)~satisfying the $4D$ 't~Hooft ansatz or its $3D$ reduction. Moreover, in this case the gauge group should be SU(2).
On the other hand, the {\it on-shell} ${\cal N}=4$ supersymmetry is compatible with the {\it general} self-dual background
and an arbitrary gauge group \cite{andr,KSm}.

Most of the  results reported here were obtained together with Francois Delduc, Sergey Fedoruk, Maxim Konyushikhin,
Olaf Lechtenfeld, Jiri Niederle\footnote{Deceased.}  and Andrei Smilga.

\section[Harmonic ${\cal N}=4$, $d=1$ superspace]{Harmonic $\boldsymbol{{\cal N}=4}$, $\boldsymbol{d=1}$ superspace}\label{section2}

As a prerequisite to the main subject, let us recall the salient features of the $d=1$ version of HSS.

\subsection[From the ordinary ${\cal N}=4$, $d=1$ superspace to the harmonic one]{From the ordinary $\boldsymbol{{\cal N}=4}$, $\boldsymbol{d=1}$ superspace to the harmonic one}

The ordinary ${\cal N}=4$, $d=1$ superspace is parametrized by the co-ordinates:
\begin{gather*}
(t, \theta^\alpha, \bar\theta_\alpha), \qquad \alpha = 1,2. 
\end{gather*}
Its harmonic extension is def\/ined as:
\begin{gather}
(t, \theta^\alpha, \bar\theta_\alpha) \quad \Rightarrow \quad (t, \theta^\alpha, \bar\theta_\alpha, u^\pm_\alpha), \quad
u^{+\alpha}u^-_\alpha = 1, \quad u^\pm_\alpha \in {\rm SU}(2)_{\rm Aut}. \label{d1harm}
\end{gather}
The main property of this $d=1$ HSS is the existence of the so-called analytic basis in it\footnote{The original parametrization (\ref{d1harm})
will be referred to as the ``central basis''.}
\begin{gather*}
  (t_A, \theta^{+}, \bar\theta^+, u^\pm_\alpha, \theta^-, \bar\theta^-) \equiv
(\zeta, u^\pm, \theta^-, \bar\theta^-) , \nonumber \\
  \theta^\pm= \theta^{\alpha}u^\pm_\alpha,   \bar\theta^\pm  = \bar\theta^\alpha
u^\pm_\alpha, \qquad  t_A = t +i(\theta^+\bar\theta^- + \theta^-\bar\theta^+) . 
\end{gather*}
Passing to the analytic basis makes manifest the presence of the analytic subspace in the $d=1$ HSS:
\begin{gather*}
(t_A, \theta^{+}, \bar\theta^+, u^\pm_\alpha)\equiv (\zeta, u^\pm) \subset (\zeta, u^\pm, \theta^-, \bar\theta^-) .
\end{gather*}
It is closed by itself under the ${\cal N}=4$ supersymmetry, but has twice as less Grassmann coordinates compared
to the full HSS. The superf\/ields given on this subspace are called {\it analytic superfields}. They can be def\/ined
by the constraints which resemble the well known chirality condition:
\begin{gather}
D^+\Phi = \bar D^+\Phi = 0 \quad \Rightarrow \quad \Phi = \Phi(\zeta, u^\pm) , \qquad
D^+ = \frac{\partial}{\partial \theta^-}, \qquad  \bar D^+ = -\frac{\partial}{\partial \bar\theta^-} .\label{analCon}
\end{gather}

An important ingredient of the HSS formalism is the harmonic derivatives, i.e.\ the derivatives with respect
to the harmonic variables:
\begin{gather*}
 D^{\pm\pm}=u^\pm_\alpha\frac{\partial}{\partial {u^\mp_\alpha}}+\theta^\pm\frac{\partial}{\partial {\theta^\mp}}
 +\bar\theta^\pm\frac{\partial}{\partial {\bar\theta^\mp}}
    +2i\theta^\pm\bar\theta^\pm \frac{\partial}{\partial {t_{\rm A}}} . 
\end{gather*}
The derivative $D^{++}$ is distinguished in that it commutes with the spinor derivatives $D^+$, $\bar D^+ $. Then, if the
superf\/ield $\Phi$ is analytic, the superf\/ield $D^{++}\Phi$ is analytic as well:
\begin{gather*}
[D^+, D^{++}] = [\bar D^+, D^{++}] = 0 \quad \Rightarrow \quad D^{++}\Phi(\zeta, u^\pm)\quad \mbox{is analytic}, \\
D^+(D^{++}\Phi) = \bar D^+(D^{++}\Phi) = 0 .
\end{gather*}
The harmonic derivatives form an SU(2) algebra:
\begin{gather*}
[D^{++}, D^{--}] = D^0, \qquad
D^0 = u^+_\alpha\frac{\partial}{\partial {u^+_\alpha}} -  u^-_\alpha\frac{\partial}{\partial {u^-_\alpha}} +
\theta^+\frac{\partial}{\partial {\theta^+}} + \bar\theta^+\frac{\partial}{\partial {\bar\theta^+}}
- \theta^-\frac{\partial}{\partial {\theta^-}} - \bar\theta^-\frac{\partial}{\partial{\bar\theta^-}} .
\end{gather*}
The operator $D^0$ counts the harmonic U(1) charge of the superf\/ields given on the $d=1$ HSS. It preserves
the analyticity and is reduced to its pure harmonic part in the central basis.

\subsection[Basic ${\cal N}=4$, $d=1$ multiplets]{Basic $\boldsymbol{{\cal N}=4}$, $\boldsymbol{d=1}$ multiplets}\label{section2.2}

It turns out that the basic of\/f-shell multiplets of ${\cal N}=4$, $d=1$ supersymmetry are represented by analytic harmonic superf\/ields subjected to
the proper additional constraints. Below we brief\/ly characterize these multiplets and quote their free superf\/ield actions. We shall use
for them the abbreviation $({\bf b, 4, 4-b})$, with ${\bf b}$ standing for the physical bosonic f\/ields and ${\bf 4 - b}$
for the auxiliary bosonic f\/ields\footnote{An additional set of admissible multiplets can be gained by changing the overall Grassmann
parity of the relevant superf\/ields.}. Depending on the choice of the action, some of the f\/ields having a~``physical'' engineering dimension can
become auxiliary, i.e.\ appear in the component action without time derivative on them.

 {\bf 1. $\boldsymbol{(4,4,0)}$ multiplet}. The multiplet $({\bf 4, 4, 0})$ is described by the superf\/ield
$q^{+ a}(\zeta, u) \propto (x^{\alpha a}, \chi^a, \bar\chi^a)$,  $a=1,2$,
with the harmonic constraint
\begin{gather}
D^{++}q^{+ a}= 0, \qquad q^{+ a} = x^{\alpha a}u^+_\alpha -2 \theta^+ \chi^a
- 2\bar\theta^+\bar\chi^a - 2i\theta^+\bar\theta^+\dot x^{\alpha a}u^-_\alpha . \label{440}
\end{gather}
The free action of this multiplet reads
\begin{gather}
S_{\rm free}(q) \sim \int dtd^4\theta du \,q^{+ a}D^{--}q^+_a \sim \int dt \left( \dot x^{\alpha a}\dot x_{\alpha a}
+ i \bar\chi^a\dot\chi_a \right). \label{440act}
\end{gather}
It also admits a Wess--Zumino (WZ) term which, in its simplest variant, is given by the analytic subspace action
\begin{gather}
S_{\rm WZ}(q) \sim \int dt_A du d\theta^+d\bar\theta^+ \,C_{(ab)}q^{+ a} q^{+ b} \sim \int dt
\, C_{(ab)} \big(i x^{\alpha a}\dot{x}_\alpha{}^b
+ 4\chi^a\bar\chi{}^b\big). \label{440wz}
\end{gather}

  {\bf 2. $\boldsymbol{(3, 4, 1)}$ multiplet}. The multiplet $({\bf 3, 4, 1})$ is described by the superf\/ield $L^{++}(\zeta, u)
\propto (\ell^{(\alpha \beta)},
 \psi^\alpha, \bar\psi^\alpha, F)$ which is subjected to the constraint
\begin{gather*}
D^{++}L^{++} = 0, L^{++} = \ell^{\alpha\beta}u^+_\alpha u^+_\beta + i(\theta^+ \chi^\alpha
+\bar\theta^+ \bar\chi^\alpha) u^+_\alpha + \theta^+\bar\theta^+ (F - 2 i\dot{\ell}^{\alpha\beta} u^+_{\alpha} u^-_{\beta}).
\end{gather*}
The free action is:
\begin{gather*}
S_{\rm free}(\ell) \sim \int dtd^4\theta du \, L^{++}(D^{--})^2L^{++} \sim \int dt \left[ \left( \dot\ell^{\alpha\beta}\dot\ell_{\alpha\beta}
-\frac{1}{2} F^2\right) +i \bar\psi^\alpha\dot\psi_\alpha\right]. 
\end{gather*}
This multiplet also admits  WZ-type ${\cal N}=4$ superf\/ield invariants.

 {\bf 3. Gauge multiplet.}  An important multiplet is the gauge one described
by an unconstrained analytic superf\/ield $V^{++}(\zeta, u)$, which exhibits
the gauge freedom with an analytic superf\/ield parameter:
\begin{gather*}
V^{++}{}' = V^{++} + D^{++}\Lambda, \qquad \Lambda = \Lambda (\zeta, u).
\end{gather*}
This gauge freedom can be f\/ixed so as to bring $V^{++}$ into the Wess--Zumino gauge with only one component $B(t)$ ($d=1$ ``gauge f\/ield''):
\begin{gather*}
V^{++}_{\rm WZ} = 2i \theta^+\bar\theta^+ B(t) , \qquad
\delta B = \dot\lambda(t).
\end{gather*}

 {\bf 4. Gauged $\boldsymbol{(4, 4, 0)}$ multiplet.} Using the superf\/ield $V^{++}$, one can def\/ine a gauged  version of the multiplet $({\bf 4, 4, 0})$.
It is represented by the superf\/ields $(v^+, \bar v^+)$, $v^+{}' = e^{i\Lambda} v^+$, $\bar v^+{}' = e^{-i\Lambda} \bar v^+$,
obeying a gauge-covariant version of the constraint in (\ref{440}):
\begin{gather*}
(D^{++} + i V^{++}) v^{+} = 0 \quad\Rightarrow \quad v^+=\phi^\alpha u^+_\alpha + \theta^+\omega_1 + \bar\theta^+\bar\omega_2
  -2i\theta^+\bar\theta^+  (\dot \phi^\alpha + i B\phi^\alpha) u^-_\alpha. 
\end{gather*}
This multiplet (``spin multiplet'') is an important ingredient of SQM models with non-Abelian background gauge f\/ields (Sections~\ref{section4} and~\ref{section5}).

 {\bf 5. Some other multiplets.} Of use in the ${\cal N}=4$ SQM model-building are also a fermionic counterpart ${\bf (0,4,4)}$
of the multiplet ${\bf (4,4, 0)}$, as well as
the multiplet ${\bf (1,4,3)}$ and the chiral multiplet ${\bf (2,4,2)}$. The f\/irst multiplet  is described by a fermionic
analog $\Psi^{+ A}(\zeta, u)$ of the superf\/ield $q^{+ a}$,
with the harmonic constraint
\begin{gather}
D^{++}\Psi^{+ A} = 0 \quad \Rightarrow \quad \Psi^{+ A} = \psi^{\alpha A}u^{+}_\alpha + \theta^+ d^A
+ \bar \theta^{+} \bar d{}^A  -2i\theta^+ \bar\theta^+ \dot{\psi}{}^{\alpha A}u^{-}_\alpha , \label{ferm1}
\end{gather}
and the free action
\begin{gather*}
S_{\rm free}(\Psi) \sim \int dt_A du d\theta^+d\bar\theta^+\, \Psi^{+ A}\Psi^+_A \sim \int dt \big(i\psi^{\alpha A}\dot{\psi}{}_{\alpha A}
- d^A\bar{d}_A\big).
\end{gather*}

The multiplets ${\bf (1,4,3)}$ and ${\bf (2,4,2)}$ can be also described within the harmonic superspace setting, though
in a rather indirect manner~\cite{Deld2,Deld3}.

Most of the analytic ${\cal N}=4$ multiplets listed here have their {\it nonlinear} counterparts, with the nonlinearly modif\/ied
harmonic constraints. Their implications in the ${\cal N}=4$ SQM models have not yet been fully explored so far. Also, in accordance
with the fact that the full automorphism group of ${\cal N}=4$, $d=1$ superalgebra is ${\rm SO}(4) \sim {\rm  SU}(2) \times {\rm SU}(2)$, each ${\cal N}=4$
supermultiplet from the above list has its ``mirror'' (or ``twisted'') counterpart, with the two SU(2) automorphism groups switching their roles.

The free actions of all these ${\cal N}=4$ multiplets can be generalized to involve a non-trivial self-interaction. The corresponding
bosonic manifolds exhibit interesting target space geometries.

\subsection{Bi-harmonic superf\/ields}\label{section2.3}

A unif\/ied description of ${\cal N}=4$ supermultiplets and their mirror cousins is achieved in the framework of {\it bi-harmonic} ${\cal N}=4$, $d=1$
HSS \cite{INi}, with the two independent sets of harmonic variables $u^{\pm 1 \alpha}$, $v^{\pm 1 i}$, $u^{1 \alpha}u^{-1}_\alpha= 1$,
$v^{1 i}v^{- 1}_i= 1$, for either two mutually commuting SU(2) automorphism groups of ${\cal N}=4$, $d=1$ supersymmetry. In this approach,
the ${\cal N}=4$, $d=1$ spinor derivatives are combined into the ${\rm SU}(2)\times {\rm  SU}(2)$ quartet,
\[
D^{\alpha i} = \left(D^\alpha, \bar D^\alpha\right),
\]
and then are split into a set of bi-harmonic projections
\begin{gather*}
D^{\alpha i}\  \Rightarrow \ \left(D^{1,1}, D^{1,-1}, D^{-1,1}, D^{-1,-1} \right), \qquad \mbox{where}\quad
D^{\pm 1 \pm 1} = D^{\alpha i}u^{\pm 1}_\alpha v^{\pm 1}_i , \quad \mbox{etc} .
\end{gather*}
In this language, the standard harmonic analytic superf\/ields discussed in the previous subsections are def\/ined by the constraints
\begin{gather}
\mbox{(a)} \ \ D^{1,1}\Phi^{(I,0)} = D^{1,-1}\Phi^{(I,0)} =0 , \qquad \mbox{(b)} \ \ D^{0,2}\Phi^{(I,0)} = 0 , \label{bih1}
\end{gather}
where $I$ is the harmonic charge with respect to the $u$-harmonics and $D^{0,2}$ is the analyticity-preserving covariant derivative with respect
to the $v$-harmonics (the harmonic constraint (\ref{bih1}b) just eliminates the $v$-dependence in the central basis). The mirror multiplets
are represented by the alternative analytic superf\/ields $\Phi^{(0,J)}$:
\begin{gather*}
\mbox{(a)} \ \ D^{1,1}\Phi^{(0,J)} = D^{-1,1}\Phi^{(0,J)} =0 , \qquad \mbox{(b)} \ \ D^{2,0}\Phi^{(0,J)} = 0 , 
\end{gather*}
with $D^{2, 0}$ being the same as $D^{++}$ def\/ined above and $J$ the harmonic charge associated with the $v$-harmonics.
These two types of ${\cal N}=4$, $d=1$ harmonic analyticity conditions cannot be imposed  on the bi-harmonic superf\/ields simultaneously,
since $\{D^{1,-1}, D^{-1,1}\} \sim \partial_t\,$.

One of the advantages of the bi-harmonic approach is that it makes manifest both SU(2) automorphism groups
in their realization on the component f\/ields. For instance, two mutually mirror ${\bf (4,4,0)}$ multiplets amount to the following sets
of $d=1$ f\/ields: $(x^{\alpha a}, \chi^{i a})$ and $(x^{i a'}, \chi^{\alpha a'})$, the f\/irst set being another form of
the multiplet~(\ref{440}). Analogously, there are two versions of the fermionic of\/f-shell multiplet ${\bf (0, 4, 4)}$,
$(\psi^{\alpha A}, d^{i A})$ and $(\psi^{i A'}, d^{\alpha A'})$.

The bi-harmonic formalism also helps to establish direct relations between various ${\cal N}=4$, $d=1$ multiplets and their twisted cousins.
Consider, e.g., a twisted version $\Psi^{(0, 1)A'}$ of the fermionic multiplet (\ref{ferm1}):
\begin{gather*}
D^{1,1}\Psi^{(0,1)A'} = D^{-1,1}\Psi^{(0,1)A'} = 0 , \qquad D^{2,0}\Psi^{(0,1)A'} = D^{0,2}\Psi^{(0,1)A'} = 0 \\
 \Rightarrow \quad
\Psi^{(0,1)A'} \propto \big(\psi^{i A'}, d^{\alpha A'}\big) .
\end{gather*}
Then the bosonic superf\/ield
\begin{gather*}
Q^{(1,0)A'} \equiv D^{1, -1}\Psi^{(0,1)A'} 
\end{gather*}
satisf\/ies the standard $u$-type harmonic analyticity constraints (\ref{analCon}), (\ref{440})
\begin{gather*}
D^{1,1} Q^{(1,0)A'} = D^{1,-1}Q^{(1,0)A'} = 0 , \qquad D^{2,0}Q^{(1,0)A'} = D^{0,2}Q^{(1,0)A'} = 0 , 
\end{gather*}
and so it is a ``composite'' version of the ${\bf (4, 4, 0)}$ multiplet,
$Q^{(1,0)A'} \propto \big(d^{\alpha A'}, \chi^{i A'} \sim \dot\psi{}^{iA'}\big)$. The invariants
(\ref{440act}) and (\ref{440wz}), upon substitution
$q^{ + a} \equiv q^{(1,0)a}  \Rightarrow   Q^{(1,0)A'}$, would produce some non-minimal actions for the multiplet $\Psi^{(0,1)A'}$ with
non-canonical numbers of time derivatives on the fermionic f\/ield $\psi^{i A'}$; in particular, the WZ-type
invariant~(\ref{440wz})  would contain
the term with two derivatives $\sim \int dt \dot{\psi}{}^{i A'}\dot{\psi}{}_{i}^{B'}C_{(A'B')}$.

For simplicity, in the subsequent consideration we shall stick to the standard ${\cal N}=4$, $d=1$ HSS with one set
of harmonic variables $u^{\pm \alpha}$.

\section[Gauging in ${\cal N}=4$, $d=1$ HSS]{Gauging in $\boldsymbol{{\cal N}=4}$, $\boldsymbol{d=1}$ HSS}\label{section3}

The ${\cal N}=4$, $d=1$ superf\/ield gauging procedure has been worked out in \cite{Deld1,Deld2,Deld3}.
It allows one to relate various ${\cal N}=4$ multiplets and their invariant superf\/ield actions with preserving, at each step,
the manifest ${\cal N}=4$, $d=1$ supersymmetry.

\subsection[A simple example of $d=1$ gauging in bosonic system]{A simple example of $\boldsymbol{d=1}$ gauging in bosonic system}\label{section3.1}

Let us start with a simple clarifying bosonic example. Consider a complex $d=1$ f\/ield $z(t)$, $\bar z(t)$ with the following Lagrangian:
\begin{gather}
L_0 = \dot{z} \dot{\bar z} +i\kappa \left(\dot z \bar z -  z\dot{\bar z}\right). \label{ex1}
\end{gather}
The f\/irst term is the kinetic energy, the second one is the simplest $d=1$ WZ term. One of the symmetries of this system is
the invariance under U(1) transformations:
\[
z' = e^{-i\lambda} z , \qquad \bar{z}' = e^{i\lambda} \bar{z} .
\]
Now we gauge this symmetry by promoting $\lambda \rightarrow \lambda(t)$. The gauge invariant action involves
the $d=1$ gauge f\/ield $A(t)$
\begin{gather*}
L_{\rm gauge} = ( \dot{z} + i A z) (\dot{\bar z} - i A \bar z) + i\kappa \left(\dot z \bar z -  z\dot{\bar z} + 2i A z\bar z\right) + 2c  A , \qquad
A' = A + \dot\lambda ,
\end{gather*}
where a ``Fayet--Iliopoulos term'' $\sim c$ has been also added. This term is gauge invariant (up to a total derivative) by itself.

The next step is to choose the appropriate gauge in $L_{\rm gauge}$:
\[
z = \bar z \equiv \rho(t) .
\]
We substitute it into $L_{\rm gauge}$ and obtain:
\begin{gather*}
L_{\rm gauge} = ( \dot{\rho} + i A \rho)\,(\dot{\rho} - i A \rho) + 2i\kappa A \rho^2 + 2c  A =
(\dot\rho)^2 + A^2 \rho^2 - 2\kappa A\rho^2 + 2c A . 
\end{gather*}
The f\/ield $A(t)$ is the typical example of auxiliary f\/ield: it can be eliminated by its algebraic equation of motion:
\[
\delta A: \ \ A = \kappa  - \frac{c}{\rho^2} .
\]
The f\/inal form of the gauge-f\/ixed action is as follows
\begin{gather}
L_{\rm gauge} \ \Rightarrow \ (\dot\rho)^2 - \left(\kappa \rho - \frac{c}{\rho}\right)^2. \label{ex3}
\end{gather}
This is a one-particle prototype of the renowned Calogero--Moser system. At $\kappa =0$, one recovers the standard conformal mechanics:
\[
L_{\rm gauge}^{(\kappa =0)} = (\dot\rho)^2 - \frac{c^2}{\rho^2} .
\]

This gauging procedure can be interpreted as an of\/f-shell Lagrangian analog of the well known Hamiltonian reduction.  In the present case,
in the parametrization $z = \rho e^{i \varphi}$, the Hamiltonian reduction consists in imposing the constraints
$p_\varphi - 2c \approx 0$, $\varphi \approx 0$, upon which the Hamiltonian of the system (\ref{ex1}) is reduced to the Hamiltonian
of (\ref{ex3}).

\subsection[An example of supersymmetric gauging in ${\cal N}=4$, $d=1$ HSS]{An example of supersymmetric gauging in $\boldsymbol{{\cal N}=4}$, $\boldsymbol{d=1}$ HSS}\label{section3.2}

Now we start from the free action of the multiplet $({\bf 4, 4, 0})$,
\begin{gather}
S = \int dt d^4\theta du\, q^{+ a}D^{--}q_a^+.\label{exa1}
\end{gather}
It is invariant under the shifts $q^{+ a} \rightarrow q^{+ a} + \lambda u^{+ a}$, $a= 1, 2$.
The gauging of this Abelian symmetry is accomplished by replacing $\lambda \rightarrow \Lambda(\zeta, u)$ and properly
covariantizing (\ref{440})
\begin{gather*}
  D^{++} q^{+ a} = 0 \quad \Rightarrow \quad \nabla^{++}q^{+ a} =  D^{++}q^{+ a} - V^{++} u^{+ a} = 0, \nonumber \\
  S \ \Rightarrow \ S_g = \int dt d^4\theta du \, q^{+ a}\nabla^{--}q_a^+, \qquad \nabla^{--}q_a^+ = D^{--}q^{+ a} - V^{--} u^{+ a}, \nonumber \\
 [\nabla^{++}, \nabla^{--}] = D^0 \quad \Rightarrow \quad D^{++}V^{--} - D^{--}V^{++} = 0, \qquad V^{--} = V^{--}(V^{++}, u). \nonumber
\end{gather*}
As the next step, we choose the gauge $u^{-a}q^+_a = 0  \Rightarrow   q^{+a} = u^{-a} L^{++}$. Then
\begin{gather*}
 D^{++}q^{+ a} - V^{++} u^{+ a} = 0\quad \Rightarrow \quad V^{++} = L^{++}, \qquad D^{++}L^{++} = 0 , \nonumber \\
  D^{++}V^{--} - D^{--}L^{++} = 0\quad \Rightarrow \quad  V^{--} = \frac{1}{2}(D^{--})^2 L^{++} , \nonumber \\
 S_g = \int dt d^4\theta du\, V^{--}L^{++} = \frac{1}{2} \int dt d^4\theta du \,L^{++}(D^{--})^2 L^{++} .
\end{gather*}
Thus, starting from the free action of the ${\bf (4,4,0)}$ multiplet and gauging a symmetry of this action, we
have eventually arrived at the free action of the multiplet $({\bf 3, 4, 1})$! As distinct from the previous example, in the present case the
gauging procedure does not produce any interaction of the multiplet $({\bf 3, 4, 1})$. Such an interaction could be induced~\cite{Deld1} if we
would gauge another Abelian symmetry of the action~(\ref{exa1}), that with respect to the U(1) transformations
$\delta q^{+a} = \lambda  C^a_b q^{+ b}$, where $C^a_b$ is a constant traceless matrix, $C^a_a = 0$.

\subsection{Further gaugings}\label{section3.3}

The superf\/ield gauging procedure just described can be equally applied to other ${\bf (4,4,0)}$ Lagrangians exhibiting some isometries and
involving an interaction, equally as to other ${\cal N}=4$, $d=1$ multiplets.
These multiplets and their superf\/ield actions can be reproduced as the appropriate gaugings  of the multiplet $({\bf 4, 4, 0})$
or of some nonlinear generalizations of the latter. Below we give a list of such gaugings:
\begin{itemize}\itemsep=0pt
\item  $({\bf 4, 4, 0})$  $\Rightarrow $ the linear $({\bf 3, 4, 1})$ multiplet -- via gauging of shifting
or rotational U(1) symmetry of $q^{+ a}$;

\item   $({\bf 4, 4, 0})$  $\Rightarrow $ the non-linear $({\bf 3, 4, 1})$ multiplet -- via gauging target space scaling symmetry,
$q^{+ a}{}' = \lambda  q^{+ a}$;

\item  $({\bf 4, 4, 0})$ $\Rightarrow$ the chiral  $({\bf 2, 4, 2})$ multiplets -- via gauging some two-generator solvable symmetries realized  on
$q^{+ a}$;

\item $({\bf 4, 4, 0})$ $\Rightarrow$  the $({\bf 1, 4, 3})$ multiplet  -- via gauging SU(2)$_{PG}$ symmetry, $q^{+ a}{}' = \lambda^a_b  q^{+ b}$;

\item  $({\bf 4, 4, 0})$ $\Rightarrow$ the fermionic $({\bf 0, 4, 4})$ multiplet  -- via gauging the semi-direct product of SU(2)$_{\rm PG}$ and the shift
symmetry $\delta q^{+ a} = \lambda u^{+ a}$.
\end{itemize}

It is worth noting that the $d=1$ gauging procedure outlined here resembles the gauging of isometries by non-propagating gauge f\/ields
in $d=2$ sigma models, which provides a f\/ield-theoretical realization of $T$-duality \cite{rv}. There is an essential dif\/ference between
the $d=1$ and $d=2$ cases, however.  An important part of the $d=2$ procedure is the insertion into the action, with a Lagrange multiplier,
the condition that the corresponding gauge f\/ield strength is vanishing. No gauge f\/ield strength can be def\/ined in $d=1$, so no analogous constraint
is possible. The gauge f\/ield f\/inally becomes just the auxiliary f\/ield of another ${\cal N}=4$, $d=1$ multiplet, and its actual role is to produce
some new potential terms in the on-shell action of the latter.

\section[${\cal N}=4$, $4D$ SQM models in the gauge field backgrounds]{$\boldsymbol{{\cal N}=4}$, $\boldsymbol{4D}$ SQM models in the gauge f\/ield backgrounds}\label{section4}

\subsection[${\cal N}=4$, $4D$ SQM with Abelian external gauge field]{$\boldsymbol{{\cal N}=4}$, $\boldsymbol{4D}$ SQM with Abelian external gauge f\/ield}\label{section4.1}

${\cal N}=4$ SQM model with coupling of $({\bf 4, 4, 0})$ multiplet to the background Abelian gauge f\/ield is described by the action \cite{hss1}:
\begin{gather}
S = \int dt d^4\theta du\, R_{\rm kin}(q^{+ a}, D^{--}q^{+ b}, u) + \int dt_A du d\theta^+
d\bar{\theta}{}^+ \,{\cal L}^{+2}(q^{+ a}, u) \equiv S_1 + S_2.\label{abel}
\end{gather}
The second term in (\ref{abel}) involves only one time derivative on the bosonic f\/ield $x^{\alpha a}$, and so it is an example of $d=1$ WZ term
\begin{gather*}
S_2 \sim \int dt \big( {\cal A}_{\alpha b}(x)\dot{x}^{\alpha b} + {\rm fermions} \big), \qquad   {\cal A}_{\alpha b}(x) =
\int du\, u^-_\alpha \frac{\partial {\cal L}^{+2}}{\partial q^{+ b}}\Big|_{\theta = 0},
\\
{\cal F}_{\alpha b\, \beta d} = \partial_{\alpha b}{\cal A}_{\beta d} - \partial_{\beta d}{\cal A}_{\alpha b}
= \epsilon_{\alpha \beta}{\cal F}_{(bd)}, \qquad {\cal F}_{(\alpha \beta)} = 0  \quad (\mbox{self-duality condition}).
\end{gather*}
Thus ${\cal N}=4$ supersymmetry requires the external gauge f\/ield to be {\it self-dual}\footnote{The analytic function
${\cal L}^{+2}(x^{\alpha a}u^+_\alpha, u^\pm)$ (prepotential) yields the {\it most general} solution of the ${\mathbb R}^4$ self-duality
constraint in the Abelian case~\cite{hk1,book}.}.  No such a~requirement is implied, e.g.,
by ${\cal N}=2$, $d=1$ supersymmetry.

How to extend this to the most interesting non-Abelian case?

\subsection{Non-Abelian self-dual background}\label{section4.2}

The coupling to non-Abelian backgrounds can be accomplished by adding the ``spin'' multiplet
$({\bf 4, 4, 0})$ \cite{IKSm}. The relevant superf\/ield action consists of the three pieces:
\begin{gather*}
S  =  \int dt d^4\theta du\, R_{\rm kin}(q^{+ a}, D^{--}q^{+ b}, u) -i\frac{k}{2}\int dt_A du d\theta^+ d\bar{\theta}{}^+ \,V^{++} \nonumber \\
\phantom{S=}{} -  \frac{1}{2} \int dt_A du d\theta^+ d\bar{\theta}{}^+ \,K(q^{+ a}, u)v^+\bar v^+
\equiv S_1 + S_2 + S_3 ,
\end{gather*}
where the new spin multiplet superf\/ields obey the constraints:
\begin{gather*}
(D^{++} + i V^{++}) v^+ = (D^{++} - i V^{++}) \bar v^+ = 0 . 
\end{gather*}

In the total action, the piece $S_1$ describes a sigma-model type interaction of $x^{\alpha a}$:
\begin{gather*}
S_1 \sim \int dt \left( f^{-2}(x)\dot x^{\alpha a}\dot x_{\alpha a} + {\rm fermions} \right),
\qquad f^{-2}(x) \sim \int du \,\Box R_{\rm kin}|_{\theta = 0} - \mbox{conformal factor}.
\end{gather*}
The term $S_2$ is the one-dimensional ``{\it Fayet--Iliopoulos}'' term:
\[
S_2 = k \int dt\, B .
\]
The term $S_3$ is most important for our purpose. It is a generalized WZ term:
\begin{gather*}
S_3 \sim \int dt \left[i \bar \varphi^\alpha ({\dot \varphi}_\alpha + iB \varphi_\alpha)
 -  \frac 12 \bar \varphi^\beta \varphi_\gamma
\left({\cal A}_{\alpha a} \right)_{\!\beta}^{\,\,\gamma} {\dot x}^{\alpha a}
+ \mbox{fermions without}\  \partial_t  \right],
\end{gather*}
where
\begin{gather*}
\left({\cal A}_{\alpha b} \right)_{\!\beta}^{\,\,\gamma}
=\frac{i}{h}
\left(\varepsilon_{\alpha\beta} \, \partial^\gamma_{\ b} h - \frac 12
 \delta^\gamma_\beta \,\partial_{\alpha b}h  \right), \qquad h(x)
 = \int du\, K(x^{+ a}, u^\pm_\beta ),\qquad \Box  h(x) = 0 .
\end{gather*}

 The background gauge f\/ield ${\cal A}_{\alpha b}$ is self-dual, ${\cal F}_{\alpha \beta} = 0$. It precisely
 matches with the general {\it 't Hooft} ansatz for $4D$ self-dual SU(2) gauge f\/ields:
\begin{gather*}
({\cal A}_{\alpha b})^\beta_\gamma \quad \Rightarrow \quad  ({\cal A}_\mu)^\beta_\gamma = \frac 12{\cal A}_\mu^i (\sigma_i)^\beta_\gamma,
\\
{\cal A}_\mu^i =  -\bar \eta^i_{\mu\nu} \partial_\nu \ln h(x) , \qquad \bar\eta^k_{ij} = \varepsilon_{kij},\qquad
\bar\eta^k_{0i} =  -\bar\eta^k_{i0} =  \delta_{ki},\qquad   i,j,k=1,2,3 .
\end{gather*}

An instructive example is the one-instanton conf\/iguration on $S^4$:
\begin{gather*}
 ds^2   = \frac {4R^4 dx_\mu^2}{(x^2 + R^2)^2} , \qquad
 {\cal A}_\mu^i   =  \frac{2R^2 \bar\eta^i_{\mu\nu} x_\nu }{x^2(x^2+ R^2)} .
\end{gather*}
It corresponds to the following choice of the functions $K(x^{+ a}, u)$ and $h(x)$:
\begin{gather*}
K(x^{+ a},u^\pm_\beta) =1+\frac{1}{\left(c^-_{a}x^{+ a}\right)^2}  , \qquad
h(x) =1+\frac{R^2}{ x_\mu^2}, \qquad c^{- a} = c^{\alpha a}u^-_\alpha, \qquad R^2 = |c|^{-2} ,
\end{gather*}
and can be brought in the BPST form, $\hat{\cal A}_\mu^i=\frac{2\eta_{\mu\nu}^i x_\nu }{x^2+R^2}$, $\hat{\cal F}^i_{\mu\nu}= - \frac {4R^2\eta^i_{\mu\nu}}{(x^2 + R^2)^2}$, by the singular gauge transformation
\begin{gather}
 {\cal A}_\mu \rightarrow \hat{\cal A}_\mu = U^\dagger {\cal A}_\mu U+iU^\dagger\partial_\mu U ,
 \qquad U(x)=-i\sigma_\mu x_\mu/\sqrt{x^2} .\label{Singg}
\end{gather}

\subsection[${\cal N}=4$  SQM with Yang monopole]{$\boldsymbol{{\cal N}=4}$  SQM with Yang monopole}\label{section4.3}

As a by-product, our non-Abelian SQM construction solves the long-lasting problem of setting up ${\cal N}=4$ SQM model
with {\it Yang} monopole as a background.

Let us consider the following bosonic Lagrangian:
\begin{gather}
L_{\,{\mathbb R}^5} = \frac 12\left(\dot{y}_5 \dot{y}_5 +  \dot{y}_\mu \dot{y}_\mu\right) + {\cal B}_\mu^{\, i} (y) \frac 12
(\bar\varphi \sigma^i \varphi)\,\dot y_\mu, \qquad \mu = 1,2,3,4 , \label{Yang}
\end{gather}
where
\[
{\cal B}_\mu^{\,i} = \frac{\eta_{\mu\nu}^i y_\nu }{r(r + y_5)} , \qquad r = \sqrt{y_5^2 + y^2_\mu} ,
\]
is the standard form of the Yang monopole potential \cite{Yang1}. Thus (\ref{Yang}) describes a coupling of
the non-relativistic particle $(y_5, y_\mu)$ in the 5-dimensional Euclidean space ${\mathbb R}^5$ to the external Yang monopole f\/ield.

After the polar decomposition of ${\mathbb R}^5$ into the angular ${\mathbb S}^4 \sim \{\tilde{y}_\mu\}$
and the radial $r$ parts as
\[
(y_5, y_\mu)
\quad \Rightarrow \quad
\left(r, \sqrt{1 - \tilde{y}^2_\mu}, \tilde{y}_\mu\right) ,
\]
and passing to the stereographic-projection coordinates as
\[
\tilde{y}_\mu = 2  \frac{x_\mu}{ 1 + x^2} ,
\]
we obtain
\begin{gather}
L_{{\mathbb R}^5} =
\frac 12\left\{\dot{r}{}^2 + 4 r^2\frac{\dot{x}_\mu\dot{x}_\mu}{(1 + x^2)^2}
\right\}
+ \frac{2 \eta_{\mu\nu}^i x_\nu \dot{x}_\mu\,\frac 12
(\bar\varphi \sigma^i \varphi) }{1 + x^2}  . \label{Yang2}
\end{gather}
The external gauge f\/ield in this Lagrangian is just BPST instanton on $S^4$. Hence, if we set the radial coordinate $r$
in (\ref{Yang2}) equal to a constant, this Lagrangian can be extended to a particular form of the Lagrangian of ${\cal N}=4$ SQM with the
self-dual SU(2) gauge f\/ield.

Thus the $5D$ mechanics with the gauge coupling to Yang monopole and ``frozen''
radial coordinate $r$ admits an extension to ${\cal N}=4$ SQM model. The radial coordinate can presumably be described
by the ${\cal N}=4$ supermultiplet ${\bf (1,4,3)}$ which also admits a description in the ${\cal N}=4$ HSS \cite{Deld2} and so can be
properly coupled to the set of the basic analytic superf\/ields~$q^{+ a}$,~$v^+$ and~$\bar v^+$.

\subsection{Quantization of spin variables}\label{section4.4}

The (iso)spin variables $\varphi_\alpha$, $\bar\varphi^\alpha$ play the pivotal role for attaining the ${\cal N}=4$ coupling to the external
non-Abelian gauge f\/ields. Let us dwell in some detail on their role in the quantum theory.

The relevant part of the total action reads:
\begin{gather}
S  = \int dt\, \big[ i \bar \varphi^\alpha (\dot{\varphi}_\alpha + iB \varphi_\alpha)   +   kB  + {\cal A}_\mu^i T^i \,\dot x_\mu\big] ,
\qquad T^i= \frac 12 \bar\varphi^\alpha\left(\sigma^i\right)^{\,\, \beta}_{\! \alpha} \varphi_\beta , \label{phiact}
\end{gather}
where
\begin{gather}
k = \mbox{integer}. \label{integer}
\end{gather}
The condition (\ref{integer})  can be deduced from the requirement of invariance of the Euclidean path integral
under topologically non-trivial gauge transformations \cite{Poly}:
\[
B(t) \  \to \ B(t) + \dot\alpha(t), \qquad \varphi(t) \ \to \ e^{-i\alpha(t)}\varphi(t) .
\]

By varying with respect to the ``gauge f\/ield'' $B(t)$, one obtains the constraint on $\varphi$, $\bar\varphi$:
\begin{gather}
  \bar\varphi^\alpha\varphi_\alpha = k . \label{Constr1}
\end{gather}
Applying the standard Dirac quantization procedure, one is left with the commutation relations:
\begin{gather*}
[ \varphi_\alpha, \bar \varphi^\beta] = \delta_\alpha^\beta   , \qquad
[\varphi_\alpha, \varphi_\beta] = [\bar \varphi^\alpha, \bar \varphi^\beta] = 0,  \qquad \varphi_\alpha \to \partial/\partial \bar\varphi^\alpha .
\end{gather*}
After quantization, the constraint (\ref{Constr1}) becomes the condition on the wave function
\begin{gather}
\bar \varphi^\alpha \varphi_\alpha \Psi =  \bar\varphi^\alpha \frac \partial {\partial \bar\varphi^\alpha} \Psi  = k\Psi  . \label{Constr2}
\end{gather}
It restricts the wave functions to be homogeneous polynomials of $\bar\varphi^\alpha$ of degree $k$.

The bilinear combinations of the spin variables $T^i$ appearing in (\ref{phiact}), after quantization are identif\/ied as SU(2) generators:
\begin{gather*}
T^i \to T^i = \frac 12\bar\varphi^\alpha\left(\sigma^i\right)^{\,\, \beta}_{\! \alpha} \frac{\partial}{\partial \varphi^\beta},
\qquad [T^i, T^k]  =  i\varepsilon^{ikl} T^l .
\end{gather*}
Taking into account the constraint (\ref{Constr2}), one derives
\begin{gather*}
T^i T^i   =  \frac 14 \left[ (\bar\varphi^\alpha \varphi_\alpha)^2 + 2 (\bar\varphi^\alpha \varphi_\alpha) \right]
=   \frac k2 \left(\frac k2 +1\right). 
\end{gather*}

Thus $T^i$ are generators of SU(2) in the irrep of spin $k/2$.
An interesting feature is that this gauge SU(2) group is at the same time the R-symmetry group
of ${\cal N}=4$ supersymmetry\footnote{The
gauge transformation (\ref{Singg}) converts this SU(2) into another SU(2) which acts on the extra indices~$a$ of~$x^{\alpha a}$
and commutes with ${\cal N}=4$ supersymmetry.}.

\section[${\cal N}=4$, $3D$ SQM in a non-Abelian monopole background]{$\boldsymbol{{\cal N}=4}$, $\boldsymbol{3D}$ SQM in a non-Abelian monopole background}\label{section5}

\subsection{Superf\/ield action in HSS}\label{section5.1}

We can choose of\/f-shell  $({\bf 3, 4, 1})$ multiplet $L^{++}(\zeta, u)$ instead of the
$({\bf 4, 4, 0})$ one $q^{+ a}(\zeta, u)$ as the dynamical (co-ordinate) multiplet and still keep the
gauged $({\bf 4, 4, 0})$ multiplet $v^+(\zeta, u)$, $\bar v^{+}(\zeta, u)$ to represent semi-dynamical spin degrees of freedom.
This gives rise to ${\cal N}=4$, $3D$ SQM with coupling to non-Abelian $3D$ gauge background~\cite{IKon}.

The corresponding total superf\/ield action is
\begin{gather*}
S  =  \int dt d^4\theta du \, R_{\rm kin}(L^{++}, L^{+-}, L^{--}, u)-\frac {i k}2  \int dt_A du d\theta^+ d\bar{\theta}{}^+ \, V^{++} \nonumber \\
\phantom{S=}{} -  \frac{1}{2} \int dt_A du d\theta^+ d\bar\theta{}^+\, K(L^{++}, u) v^+\bar{v}{}^+ \equiv S_1 + S_2 + S_3 . 
\end{gather*}
The f\/irst two pieces produce the kinetic sigma-model type term
of the $({\bf 3, 4, 1})$ multiplet and Fayet--Iliopoulos term of the gauge ${\cal N}=4$ multiplet. The third piece describes
the WZ-type superf\/ield coupling of the co-ordinate multiplet to the external gauge background.

\subsection{Component action}\label{section5.2}

For simplicity, we choose the free action for $L^{++}$, with the Lagrangian $\sim L^{++}(D^{--})^2L^{++}$.
The component form of the bosonic part of the full action $S$ is
\[
S \to \int dt \left[\frac 12\,\dot \ell_m^2
  +{\cal A}_m^i T^i\dot \ell_m
  +i\bar\varphi^\alpha\left(\dot\varphi_\alpha+iB\varphi_\alpha\right)+kB
+\frac 18 F^2
  +\frac 12 F\left( U^i T^i\right) \right],
\]
where $i=1,2,3$, $m = 1,2,3$ and
\begin{gather*}
{\cal A}_m^i
  =-\varepsilon_{mni} \partial_n \ln h,\qquad
  U^i=-\partial_i \ln h , \qquad  T^i= \frac{1}{2}\bar\varphi^\alpha \left(\sigma^i\right)_{\!\alpha}^{\,\,\beta}\varphi_\beta ,\\
  h(\ell)=\int du\, K\big(\ell^{\alpha\beta}u^+_\alpha u^+_\beta,u^\pm_\gamma\big), \qquad \Delta h = 0 .
  \end{gather*}

The $3D$ gauge f\/ield ${\cal A}_m^i$ and potential $U^i$ are particular solutions
of the {\it Bogomolny} equations
\[
{\cal F}_{mn}^i = \varepsilon_{mns}\nabla_s U^i ,
\]
with
\begin{gather*}
{\cal F}_{mn}^i = \partial_m{\cal A}_n^i - \partial_n{\cal A}_m^i
+\varepsilon^{ikl}{\cal A}_m^k{\cal A}_n^l , \qquad \nabla_m U^i=\partial_m U^i +\varepsilon^{ikl}{\cal A}_m^k U^l .
\end{gather*}

\subsection{Quantization and SO(3) example}\label{section5.3}

Quantization follows the same line as in the $4D$ case:
\begin{gather*}
[T^i, T^k] = i\varepsilon^{ikl}T^l ,
\qquad T^iT^i = \frac{k}{2}\left(\frac{k}{2} +1 \right).
\end{gather*}

The Hamiltonian, in the case with the free kinetic term for $\ell_m$, is
\begin{gather}
H=\frac{1}{2}\left(\hat p_m-{\cal A}_m\right)^2
+\frac 12 U^2 + \mbox{fermionic  terms}, \qquad U \equiv U^iT^i . \label{Syst}
\end{gather}
A new feature of the $3D$ case is the appearance of the ``induced'' potential term $\sim U^iU^kT^i T^k$ which is generated
as a result of elimination of the auxiliary f\/ield~$F$. The system~(\ref{Syst}) provides a non-Abelian generalization of the ${\cal N}=4$  SQM
model pioneered in~\cite{Crombr}.

As an example of the gauge-f\/ield background, let us quote the SO(3) invariant one:
\begin{gather*}
h_{{\rm so}(3)}(\ell) = c_0 + c_1 \frac{1}{\sqrt{\ell^2}}\quad \Rightarrow \quad
{\cal A}^i_m= \varepsilon_{mni}\frac{\ell_n}{\ell^2}  \frac{c_1}{c_1 + c_0\sqrt{\ell^2}}, \qquad U^i =
\frac{\ell_i}{\ell^2}  \frac{c_1}{c_1 + c_0\sqrt{\ell^2}} . 
\end{gather*}
In the limit $c_0 = 0$ the background gauge f\/ield becomes {\it Wu--Yang} monopole \cite{WuYang}; the ${\cal N}=4$
SQM for this case was earlier constructed in \cite{BKS} in a dif\/ferent approach\footnote{In \cite{BKS}
and in some other works of these authors, it was suggested to describe the spin variables, originally introduced in \cite{Calog}
as a bosonic sector of the multiplet ${\bf (4,4,0)}$, by the fermionic multiplet ${\bf (0,4,4)}$. On the road from the superf\/ield
action to the component one, these authors make non-canonical replacements of the time derivatives of the fermionic f\/ields by new auxiliary
fermionic f\/ields. This procedure basically amounts to the construction of ``composite'' ${\bf (4,4,0)}$ multiplet
from the ${\bf (0,4,4)}$ one, as explained in the end of Section~\ref{section2.3}. In view of existence of the direct ${\cal N}=4$ of\/f-shell
superf\/ield formulation of the multiplet ${\bf (4,4,0)}$, the use of the auxiliary fermionic multiplet as the starting point looks
artif\/icial and superf\/luous.}.

\section{Summary and outlook}\label{section6}

Let us summarize the basic contents of this contribution.

  One of its incentives was to provide more evidence that the ${\cal N}=4$, $d=1$ harmonic superspace~\cite{hss1} is a useful tool of
  constructing and analyzing SQM models with ${\cal N}=4$ supersymmetry. It allows one to construct of\/f-shell invariant actions,
  to establish interrelations between dif\/ferent multiplets, to reveal the relevant target geometries, and so on.

  As one of the recent uses of the $d=1$ HSS, of\/f-shell ${\cal N}=4$ supersymmetric couplings of the multiplets $({\bf 4, 4, 0})$
  and $({\bf 3, 4, 1})$
  to the external non-Abelian gauge backgrounds were presented. They essentially  exploit the auxiliary (iso)spin $({\bf 4, 4, 0})$ multiplet.
  The background should be self-dual and be described by the 't Hooft ansatz or its static $3D$
  reduction. HSS is indispensable for setting up the relevant of\/f-shell actions. It should be noticed that, for the time being,
  our {\it off-shell}
  superf\/ield construction is limited to the 't Hooft ansatz  and the gauge group SU(2). It is still an open question   how to extend it
  to the SU(N) gauge group and to the general self-dual backgrounds, e.g., to the renowned ADHM one.

  Surprisingly, the {\it on-shell} actions, with all auxiliary f\/ields eliminated, admit a direct extension to the gauge group SU(N)
  and general self-dual backgrounds \cite{KSm,IKSm,IKon}. This is attainable at cost of {\it on-shell} realization of ${\cal N}=4$ supersymmetry.
  It is interesting to inquire if it is possible to derive these models from some {\it off-shell} superf\/ield approach.

We f\/inish by indicating possible applications and further directions of study.

It would be interesting to extend our construction of couplings to the external non-Abelian gauge f\/ields to the case
of higher ${\cal N}$, $d=1$ supersymmetries, e.g.\ to ${\cal N}=8$. Also, an obvious task is to exploit some
other ${\cal N}=4$ multiplets to represent the coordinate and/or spin variable sectors, e.g.\ nonlinear versions of the multiplets
$({\bf 4, 4, 0})$ and $({\bf 3, 4, 1})$, the ${\bf (2, 4, 2)}$ multiplets, etc.

As for applications, it would be tempting to make use of the techniques based on the semi-dynamical spin supermultiplets
for the explicit calculations of the world-line superextensions of non-Abelian Wilson loops, with the evolution parameter
along the loop as a ``time''. One more possible area of using the models constructed and their generalizations
includes superextensions of Landau problem and higher-dimensional quantum Hall ef\/fect, as well as supersymmetric
black-hole stuf\/f.

\subsection*{Acknowledgements}

\noindent I thank the Organizers of the Workshop ``Supersymmetric Quantum Mechanics and Spectral Design''
(Benasque, July 18--30, 2010) for inviting me to participate and for the warm hospitality in Benasque. I am grateful
to my co-authors in \cite{Calog, hss1,Deld1,Deld2,Deld3,IKSm,IKon,INi} for the fruitful collaboration.
A partial support from the RFBR grants 09-02-01209, 09-01-93107, 09-02-91349, as well as from a grant
of the Heisenberg--Landau Program, is cordially acknowledged.

\pdfbookmark[1]{References}{ref}
\LastPageEnding

\end{document}